\documentclass[aps,twocolumn,groupedaddress, showpacs,prb]{revtex4}
\usepackage{graphicx}%

\begin{document}
\bibliographystyle{apsrev}


\title{Metric-first \& entropy-first surprises}


\author{P. Fraundorf}
\affiliation{Physics \& Astronomy/Center for NanoScience, U. Missouri-StL (63121)}
\affiliation{Physics, Washington University (63110), St. Louis, MO, USA}
\email[]{pfraundorf@umsl.edu}


\date{\today}

\begin{abstract}

Established idea-sets don't update seamlessly. The tension between new and old views of nature is e.g. documented in Galileo's dialogs and now present in many fields. However the science of Bayesian model-selection has made recent strides in both life \& physical sciences, in effect suggesting that we look to models which are quantitatively {\em surprised least} by present-day observations. 

We illustrate the relevance of this to physics-education with a qualitative look at two paradigm-shifts, namely from {\bf Lorentz-transform to metric-equation} descriptions of motion in space-time, and from {\bf classical to statistical thermodynamics} with help from Boltzmann's choice-multiplicity \& Shannon's uncertainty. Connections of the latter to {\bf correlation measures} behind available-work, evolving complexity, and model-selection relevant to physics undergrads are also explored. 

New strategies are exemplified with Appendices {\em for teachers} on: anyspeed traffic-laws \& 3-vector velocity-addition, the energy-momentum half-plane lost to finite lightspeed, the modern distinction between proper \& geometric accelerations, metric-first kinematics with acceleration \& differential-aging, quantifying risk with a handful of coins, effective number of choices, available work in bits, reversible-thermalization of life's power-stream, and choice-multiplicity measures of layered complex-system health.

\end{abstract}
\pacs{05.70.Ce, 02.50.Wp, 75.10.Hk, 01.55.+b}
\maketitle

\tableofcontents

\section{Introduction}
\label{sec:Intro}

Thomas Kuhn in his book\cite{Kuhn70} on the structure of scientific revolutions illustrated, with help especially from physics examples, how the evolution of well-worn approaches naturally encounters resistance from experts in the old\cite{Planck1949}. Similarly Martin Gardner in his book on parity inversion\cite{Gardner90} cites Hermann Kolbe's negative reaction to the prediction of carbon's tetrahedral nature by Jacobus van't Hoff (Chemistry's first Nobel Laureate).  The hullabaloo\cite{Doebeli2010} about the Nowak et al. paper\cite{Nowak2010} on models for evolving insect social behavior is a more recent research example, while participants in the content-modernization branch of physics education research (PER) have engaging tales on the education side\cite{simplifications1999}.

Kuhn discusses new paradigms in terms of their ability to adapt to new observations, while continuing to fill the role of earlier models. Sometimes this is done by preserving the old models as special cases of new ones. Thus relativity preserves Galilean kinematics as limiting behavior at low-speeds, and quantum mechanics preserves Newtonian dynamics in the limit of high-mass.

The science of Bayesian inference applied to model-selection may eventually offer a quantitative way to assess new approaches. That's because log-probability based Kullback-Leibler divergence, or {\bf the extent to which a candidate model is surprised} by emerging insights, formally considers both goodness of fit (prediction quality) and algorithmic simplicity (Occam's razor). This strategy lies at the heart of {\em independent approaches} to quantitative model-selection in both the physical and the life sciences.

We review the basis for {\em quantitative} Bayesian model-selection in this paper, perhaps for the first time ``joined up'' in context of these independent developments, and then {\em qualitatively} review old news e.g. that: (i) proper velocity and acceleration are useful over a wider range of conditions than their coordinate siblings; (ii) uncertainty, and uncertainty slopes like coldness 1/kT, provide insight into a wider range of thermal behaviors than does T alone; and (iii) entropy has deep roots in Bayesian inference, the correlation measure KL-divergence, and simple measures for effective freedom of choice. A set of ten potentially-useful appendices is then provided for those looking to explore specific aspects of these connections with their class. 

\section{Principles}
\label{sec:Principles}

Bayesian approaches to quantitative model-selection, which show the assumptions behind standard statistical-recipes as sub-cases, are slowly establishing themselves in the physical and life sciences. In astronomy, physics and engineering this is happening with help from maximum-entropy approaches\cite{Gregory2005}. In behavioral ecology and the life sciences it is happening with Kullback-Leibler (KL) divergence based approaches like Akaike Information Criterion\cite{Burnham2002} (AIC). As discussed later in this paper, most senior thermal-physics texts already reflect this shift.

Log-probabilities as surprisals\cite{Tribus61} of the form $k \ln[\frac{1}{p}] \ge 0$ (where constant $k$ is $\frac{1}{\ln[2]}$ for bits, 1 for nats, etc.) add when probabilities multiply, and are especially useful in Bayesian inference when constraints on parameter-averages are operational. Hence they they are important tools not just in physics, but also in the formal extension of logic to cover any decisions based on partial information\cite{Jaynes2003}.

\subsection{least-squares model-selection}

\begin{figure}
\includegraphics[scale=0.7]{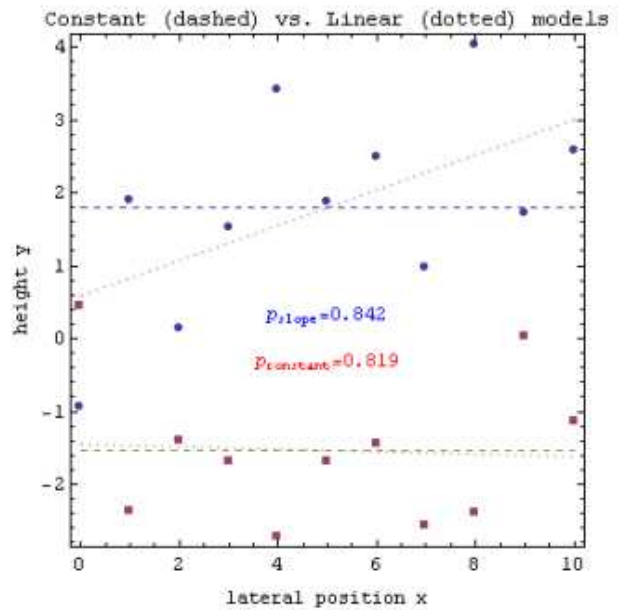}%
\caption{Probabilities of constant vs. linear fits using AIC.}
\label{linQuad}
\end{figure}

By way of concrete example, $N$ measurements of a process-parameter $y$ (with resolution $\delta y$ and normally-distributed errors having a standard-deviation of $\sigma_y$) are expected to have an uncertainty (average surprisal) in natural information units (nats) of
\begin{equation}
S_{\text{q/q}} \simeq N \ln\left[\frac{\sqrt{2 \pi}\sigma_y}{\delta y}\right] + \frac{N}{2} \ge 0 
\label{gaussianErrors}
\end{equation} 
Generally the $y$-resolution $\delta y$ of the data should be much smaller than $\sigma_y$, i.e. the measurement would hopefully have some empty resolution. If the two are equal one expects about 2.05 bits of surprisal (or 1.42 nats) {\em per data point} when one's model and reality match perfectly, which will increase as $y$-resolution becomes finer.

To a process-model $f[\hat{A},x]$ having optimized fit-parameters $\hat{A}_k$ for $k=1,K$, the average-surprisal on encountering the data is instead the log of the reciprocal likelihood, where likelihood $p[D|AM]$ is the probability of the data given a model with parameter-set $A_k$ for $k=1,K$. This is in turn equal to the log of one over the maximum-likelihood $p[D|\hat{A}M]$, plus an Occam-factor term $\ln[1/\Omega_M]$ obtained from fit-parameter marginalization that measures the amount of presumption needed to arrive at the optimized parameter-set $\hat{A}$. In other words this is the cross-entropy:
\begin{equation}
S_{\text{p/q}} \simeq \underbrace{N \ln\left[\frac{\sqrt{2 \pi}\sigma_y}{\delta y}\right] + \frac12 \sum_{i=1}^N \left(\frac{y_i-f[\hat{A},x_i]}{\sigma_y} \right)^2}_{\text{uncertainty in best-fit}} + \underbrace{\ln\left[\frac{1}{\Omega_M}\right]}_{\text{cost of fit}}
\label{surpriseAtData}
\end{equation}
so that the expected net-surprisal or KL-divergence $I_{\text{p/q}} = S_{\text{p/q}}-S_{\text{q/q}}$ between model and reality in nats is:
\begin{equation}
I_{\text{p/q}} \simeq \frac12 \sum_{i=1}^N \left(\frac{y_i-f[\hat{A},x_i]}{\sigma_y} \right)^2 + \ln\left[\frac{1}{\Omega_M}\right] - \frac{N}{2}
\label{KLdivOfData}
\end{equation}
where on average $I_{\text{p/q}} \ge 0$ is a measure of the deviation between model probabilities and the reality that generated the datapoints $\{x_i,y_i\}$ for $i=1,N$. KL-divergence will also come up here in our discussion of engineering-measures for available-work. 

\begin{table*}
\caption{Paradigms to describe the building-blocks of atoms.}
\begin{tabular}{r||c|c|c|c}
models $\mapsto$ & particle/wave & Bohr-like & momentum's link to & Feynman's view:\\
nature's surprises & dichotomy & hybridizations & spatial-frequency\cite{deBroglie1925} & explore all paths\cite{Taylor98b}\\ 
\hline \hline
atom levels are quantized & ! & & &	\\
\hline
particle-beams diffract & ! & ! & & \\
\hline
single-particles self-interfere	& !	& ! & ! &	\\
\hline
mapping deBroglie phase-shifts across single electrons\cite{Spence1988} & ! & !& ! & \\
\hline \hline
\end{tabular}
\label{table1}
\end{table*}

Although this last expression requires knowledge of the generating probability-set $q$, the resolution-dependent term in the previous expression will be the same for any model-function $f[A,x]$ and associated probabilities $p$ used to analyze the same data. Hence the smallest $S_{\text{p/q}}$ value may be sought among different models as the best bet, even if the generating probabilities $q$ are unknown.

The optimum parameters $\hat{A}_k$ are of course estimated by the usual least-squares minimization process, but the Occam factor is less familiar to physicists. Gregory\cite{Gregory2005} gives an explicit expression for it when the model is linear in the parameters, in terms of the parameter covariance-matrix and the range $\Delta A_k$ of linear prior-probabilities for the parameter values.

A more general approach to models which are clearly simplifications of a complex reality (e.g. in the behavioral sciences) involves AIC\cite{Burnham2002} which (for large $N$) estimates $\ln[1/\Omega_M]$ as approximately $K$ (the number of fit parameters). In the absence of specific information, this provides a rule for defining fit-parameter prior-probability ranges in terms of parameter covariance-matrix eigenvectors. AIC then simply compares $S_{\text{p/q}}$ values (the average surprisal of model at the data) for competing models, thus allowing for an unbiased comparison of models while making no assumptions at all about reality's uncertainties i.e. about $S_{\text{q/q}}$.

Use of AIC to decide if two sets of data points are better modeled with a sloped-line ($K=2$), or with a constant ($K=1$), is illustrated in Fig. \ref{linQuad}. To determine each model's probability $p$, they are given equal priors thus setting $p$ proportional to the marginalized-likelihoods $e^{-S_{\text{p/q}}}$, and then normalized to $1$. Like other goodness-of-fit measures (which can be seen as special cases of this) the Occam factor term in such surprisal analyses in effect adds a penalty for superfluous fit-parameters. 

\begin{figure}
\includegraphics[scale=0.68]{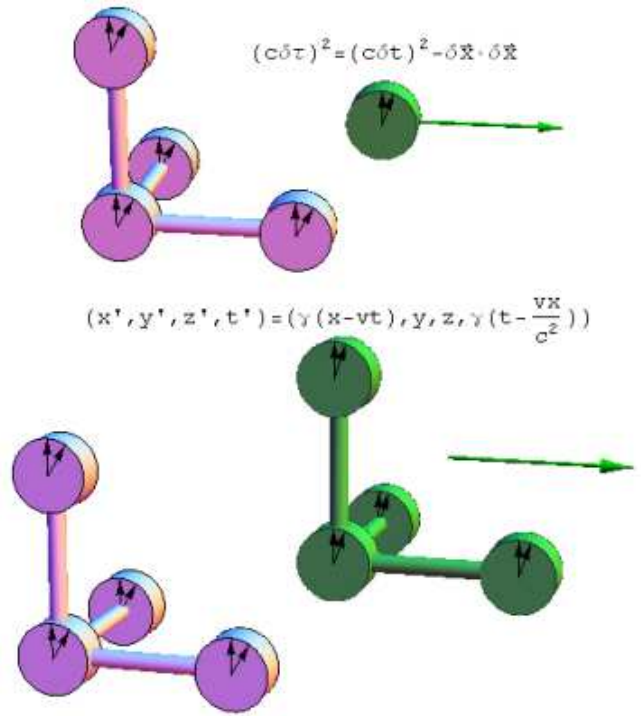}%
\caption{Motion described with one, or two, map-frames of yardsticks \& synchronized clocks.}
\label{metricLorentz}
\end{figure}

\subsection{classroom model-analysis}

Although we are not ready here to similarly {\em quantify} evolving classroom paradigms, we propose to {\em qualitatively} look (e.g. through simple tables) at the extent to which various models of the physical world have been surprised by newly-recognized phenomena. We illustrate this with a few examples, based on content changes already underway in the evolving physics curriculum. In the process we'll get a chance to further explore the curious relationship between the technical concept of surprisal, and vernacular uses for the word surprise.

A sample table on paradigms for the building blocks of atoms, a subject not discussed in this paper, is provided in Table \ref{table1}. Exclamation marks indicate that a given model (column) was (at least at first glance) a bit surprised by a given phenomenon (row), even if with added work it could accomodate the effect. Similar tables for your own perspective on phenomena and models covered in any given class, as it relates to the textbook in hand as well as the larger picture, may be worth putting together for sharing with your students and (e.g. in electronic collaboration spaces) with the larger teaching community as well. 

\section{Metric-based motion}
\label{sec:Motion}

As shown at the bottom of Fig. \ref{metricLorentz}, the traditional path to special relativity involves a pair of co-moving map-frames with their own yardsticks and synchronized clocks. A bit of experimentation with x-ct diagrams, for example, will show that these dual frames may be necessary for a complete look at the effects of length-contraction and frame-dependent simultaneity.

However the figure illustrates also how all of the variables in the metric equation may be examined in detail using only one map-frame (with yardsticks and synchronized-clocks), along with a traveling clock. This is important because the metric equation serves up a more compact summary of relativistic kinematics, and also paves the way to acceleration and gravity curved space-time.

Einstein's own path to relativity\cite{Einstein96,Pais82} began with dual frames and an impression of Minkowski's tensor-form as `\"{u}berfl\"{u}ssige Gelehrsamkeit' (superfluous erudition), before recognizing the metric as a key to understanding both accelerated-frames and gravitation. The goal here is to explore advantages (and challenges for students) of starting with the metric equation, before addressing problems that require comoving systems of yardsticks and synchronized-clocks.

In this context, Table \ref{table2} has columns for various historical-approaches to describing motion and its causes. The rows are for select observational phenomena that helped to direct our choice of model from one historical-approach to the next. 

\begin{table*}
\caption{Paradigms describing macroscopic motion and its causes.}
\begin{tabular}{r||c|c|c|c}
models $\mapsto$ & Aristotelian & Galilean & Lorentz-1st & metric-1st\\
nature's surprises & dynamics & Newtonian & relativity\cite{Einstein20} & dynamics\cite{Minkowski08}\\ 
\hline \hline
force-free coasting & ! & & &	\\
\hline
frame-dependent clocks & ! & ! & & \\
\hline
the lightspeed curtain	& !	& ! &  &	\\
\hline
single map-frame analyses & ! & !& ! & \\
\hline
three-vector velocity-addition & ! & !& ! & \\
\hline
proper vs geometric accelerations & ! & !& ! & \\
\hline
accelerated-frame curvature\cite{DolbyGull01} & ! & !& ! & \\
\hline
gravitationally-curved spacetime\cite{Einstein1915} & ! & !& ! & \\
\hline \hline
\end{tabular}
\label{table2}
\end{table*}

The transition from first to second columns (Aristotelian to Galilean/Newtonian dynamics) was e.g. the subject of Galileo's ``Dialog concerning the two chief world systems'', written as a play\cite{Galileo62} to keep him from getting in trouble with the church. Galileo, trained to analyze motions in terms of a late medieval impetus-theory which held that continued motion of an object results from an internal power implanted in it by its projector\cite{Kuhn70}, turned his quantitative observational-focus toward the concept of acceleration in effect coming up with our familiar equations of constant acceleration. This set the stage for Newton's formulation of the principle of inertia, which in effect made unchanging speed the default rather than a process in need of explanation.

The transition from the second to third columns (Newtonian to relativistic dynamics) was driven by the connection between electricity and magnetism\cite{Einstein05}, before we actually had experimental data e.g. on velocity-dependent (and on gravitational) time-dilation. Nonetheless the inferred upper-limit on coordinate-speed, plus the speed \& location dependence of time's passage, persist today as a source of surprise to beginners growing up in Newtonian worlds.

The transition from the third to fourth columns (from transformation-first to metric-first approaches) is more subtle since the two share common axiomatic roots. As Einstein recognized in trying to incorporate gravity, however, the difference is not superfluous. In fact introphysics textbooks increasingly show benefits of the metric-first approach\cite{Moore2003} e.g. via concepts of proper-time and proper-length used to break the {\em symmetry} between dual-frames which was a historical trademark of the Lorentz-first tradition. Since momentum is simply rest-mass times the proper-velocity vector at all speeds\cite{SearsBrehme68,Shurcliff96}, there's no need for a speed-dependent mass\cite{Adler87,Okun2009,Sandin1991,Oas2005} to preserve that relationship.

The first surprise for intro-physics class participants may be that insight into motion at high speeds does not have to begin with the dreaded equations of relative motion along with consideration of {\bf two} separate-sets of yardsticks, synchronized-clocks, and definitions of simultaneity. The metric-first approach allows one to get quite a lot done in the context of a single reference ``map-frame'' of yardsticks and synchronized clocks, and therefore a single reference-definition of simultaneity between events. The main new thing for Newtonians is this: Time-elapsed on a clock depends on its motion with respect to the reference map.

The second surprise for Lorentz-first proponents may be that proper-velocity (map-distance traveled per unit travler-time i.e. $\vec{w} \equiv d\vec{x}/d\tau = \gamma \vec{v}$) is the most robust\cite{SearsBrehme68,Shurcliff96} of several velocity-parameters that arrive with the metric-equation's frame-invariant proper-time variable $\tau$. Two other velocity-parameters (not including rapidity) are our old friend coordinate-velocity $\vec{v} \equiv d\vec{x}/dt$, and the scalar Lorentz-factor $\gamma \equiv dt/d\tau$. Proper-velocity and coordinate-velocity are the same at low speeds, but (as we show in Appendices \ref{appxA} and \ref{appxB}) it is proper-velocity that is most directly-connected to traffic-safety, and which retains the property of 3-vector additivity at high speeds.

In Appendix \ref{appxC}, proper-velocity's 4-vector partnership with Lorentz-factor and its role as momentum per unit mass yields a view of the kinetic-energy/momentum continuum that might surprise Newton (and those like Maxwell who fleshed in the relevance of his ideas to energy) as well. The finite speed of light does not put an upper limit on proper-velocity like that on coordinate-velocity, but it does drop a curtain on the free-particle dispersion relation i.e. the range of allowed kinetic-energy/momentum pairs.

The metric-first approach, perhaps surprisingly, does not have the same reservations as does the Lorentz-first approach about addressing accelerated-travel\cite{French68,Cheng2005}. One benefit of this (discussed in Appendix \ref{appxD}) is an option at the outset to adopt the equivalence-principle dichotomy of accelerations and forces as either proper or geometric, while at the same time distinguishing ``traveler-felt'' proper-accelerations from that coordinate-acceleration which (like coordinate-velocity) equals the proper-quantity only at low speeds. Those geometric ``affine-connection'' effects, locally explainable in terms of ``forces'' that act on every ounce of an object but vanish with suitable choice of a free-float reference frame, include familiar inertial-forces and may also include gravity.

A more immediately-practical result is that anyspeed-equations for constant proper-acceleration follow which are direct analogs of the low-speed equations of constant acceleration. These are discussed in Appendix \ref{appxE}. If this ability to track accelerated motion is combined with the radar-time simultaneity definition of Dolby and Gull\cite{DolbyGull01}, this also opens the door to calculus-only exploration of accelerated-frame curvature in flat space-time.

Finally, of course, the metric equation allows one to explain gravity's action at a distance. The recent book by Taylor and Wheeler\cite{Taylor01} on scouting black holes with calculus shows how only minor changes to the flat-space metric equation allows introductory physics students to have fun with high-curvature environments well. 

\section{Multiplicity-based thermodynamics}
\label{sec:ChoiceMultiplicity}

\begin{table*}
\caption{Paradigms describing order-disorder processes in complex systems.}
\begin{tabular}{r||c|c|c|c}
models $\mapsto$ & Rennaisance & early 19th century & entropy-first & correlation-first\cite{ShannonWeaver49,Jaynes57a}\\
nature's surprises & engineering & thermodynamics & inference & inference\\ 
\hline \hline
heat's mechanical equivalent & ! & & &	\\
\hline
entropy $\Leftrightarrow$ accessible-states: $S = k_{\text{Boltzmann}} \ln[W]$ & ! & ! & & \\
\hline
Gibbs' ensemble constraints	& !	& ! &  &	\\
\hline
Onsager reciprocity \& fluctuations	& !	& ! &  &	\\
\hline
communication-theory applications & ! & ! &  & \\
\hline
multiplicity $\Rightarrow PV=NkT, E/N=(\nu/2)kT$ etc. & ! & ! &  & \\
\hline
negative absolute reciprocal-temperatures\cite{Castle65} & ! & ! &  & \\
\hline
negentropy\cite{Brilloun62} \& reversible computation & ! & ! &  & \\
\hline
KL-divergence as available-work\cite{Tribus71,Lloyd89b} & ! & ! & ! & \\
\hline
correlation-analysis of molecule-codes, neural-nets, etc. & ! & ! & ! & \\
\hline
evolution of multi-layer complexity & ! & ! & ! & \\
\hline \hline
\end{tabular}
\label{table3}
\end{table*}

Another well-defined paradigm-shift in the undergraduate physics curriculum is the transition from historical-axiomatic, to statistical, approaches to thermal physics. In fact since the middle of last century, many of the senior undergraduate texts have taken the statistical path including those by Kittel \& Kroemer\cite{Kittel80p}, Keith Stowe\cite{Stowe84p}, Dan Schroeder\cite{Schroeder00}, and Claude Garrod\cite{Garrod95} who refers to reciprocal-temperature as coldness. 

We discuss this shift here for two reasons. The first is that the transition has not been exploited in many introductory physics texts\cite{Moore97}, with one exception being Tom Moore's introductory physics Unit T\cite{Moore2003}.

The second reason is that the tools of statistical inference behind this transition in thermal physics are relevant to a much wider range of complex-system applications to which physics courses can make a contribution. One of these, in fact, is the science of model-selection whose principles were discussed at the beginning of this paper.

The big surprise for first-column theories e.g. of phlogiston and caloric\cite{Kuhn70} may have been Joule's observations of heat's mechanical-equivalent. Phenomenological discoveries of not yet well-explained gas laws and calorimetric behaviors were also taking place in this time-frame.

Entropy-first inference had its roots in the late 19th century, e.g. with the ensemble work of J. W. Gibbs and the kinetic-theory work of L. Boltzmann. However its recognition as a branch of statistical inference in general had to await its application to communication systems by C. Shannon\cite{ShannonWeaver49} and the joining up of the physics and statistical-inference threads by E. T. Jaynes\cite{Jaynes57a,Jaynes57b,Jaynes1963} and colleagues.

From the pedagogical point of view, the key strength of the entropy-first approach is that it provides physical-model assumptions to describe when the ideal-gas law (e.g. accessible states $W$ proportional to $V^N$), equipartition (e.g. accessible states $W$ proportional to $E^{\nu N/2}$), and mass action will serve as good approximations in place of the phenomenological and/or axiomatic assertions available before that. Senior undergraduate texts already reflect these insights.

Another surprise to the early 19th century view was the utility of negative absolute-temperatures in sub-systems with an upper limit on total energy. In spite of a lovely book on this subject by Westinghouse engineers\cite{Castle65} in the early 1960's, the fact that energy's uncertainty slope (i.e. reciprocal-temperature or coldness between plus and minus infinity) drives the flow of heat from low to high slope remains poorly-explained by scientists\cite{Braun2013} and a mystery to lay-persons (and some meteorologists) today.

\begin{figure}
\includegraphics[scale=0.60]{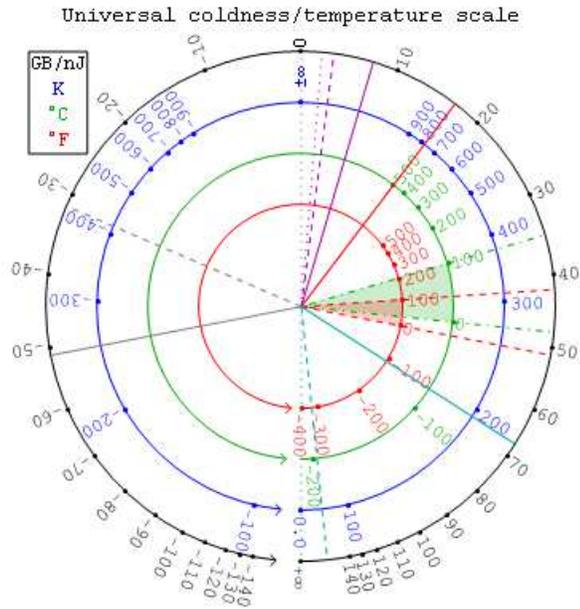}%
\caption{Coldness (1/kT) scale with nuclear/atomic inverted-population states on left \& heat-flow from low to high.}
\label{coldnessScale}
\end{figure}

A related suprise to 19th century thermodynamicists may have also been (since $1/kT = dS/dE$) that the natural (as distinct from historical) units for temperature become energy per unit information, and for heat capacities become bits\cite{pf.hcapbit}. In these contexts a universal coldness/temperature-scale like that in Fig. \ref{coldnessScale} might help your students move past the problem quickly. 

The uncertainty-slope is measured in gigabytes per nanoJoule because these are everyday concepts, although the value is only about 9/8ths of $1/kT$ in nats per electronVolt so that the familiar room-temperature value for kT as $1/40$ eV places room temperature on this scale naturally around $40$ as well. Radial lines mark benchmark temperatures in a variety of physical systems, including selected atomic and nuclear population inversions. As an added advantage, the plot maps the infinite reciprocal-temperature manifold onto a finite interval using the $\ln[N/n-1]$ approximation to a spin-system's entropy derivative.

However the transition from entropy-first to correlation-first is less well-discussed in texts, even though it has wide-ranging cross-disciplinary applications to which physics-students can contribute. One of these is the science of surprisal-based model selection being explored qualitatively here. We expand a bit, therefore, on that transition in the last section of this paper.

\section{Correlation-first inference}
\label{sec:MatchupMultiplicity}


To explore the connection between correlation-first and entropy-first inference, we step back from entropy and multiplicity to probability measures and then forward to correlation measures. One advantage of this approach is that it starts and ends with probabilities and statistical inference, thus bypassing questions about the rationale for its use in equilibrium thermodynamics\cite{Jaynes1963}. Such tools of Bayesian inference might in fact be applied to any situations about which one has only partial information.

Telling students about the connection between entropy-first and correlations-first approaches will also allow physics to make contact with a number of other lively disciplines, some of which are touched on in our appendices. Non-physics majors, in particular, may never hear about these connections if they aren't mentioned in at least one of their physics classes.

\subsection{surprisals}
\label{sec:surprisals}

Recall that information units can be introduced by the statement that $\#$ choices equals $2^{\#\text{bits}}$ or $e^{\#\text{nats}}$.  In this way very small probabilities p can be put into everyday terms as the {\bf surprisal}\cite{Tribus61} in bits of tossing n coins all heads up, since $p = 1/2^{\#\text{bits}}$, with the added advantage that surprisals add whenever their probabilities multiply (Appendix \ref{appxE}). Evidence in bits\cite{Jaynes2003} for a true-false proposition can similarly be written as $e[p] = s[1-p]-s[p]$, where surprisal is $s[p]=\ln_2[1/p]$.

All of these applications rely on the fact that probabilities between 0 and 1 can be written as multiplicities $w_p = 1/p$ between 1 and $+\infty$ or as surprisals between 0 and $+\infty$ using information units determined by the constant k in the expression $s_p = k \ln[1/p]$. This surprisal $\Leftrightarrow$ multiplicity $\Leftrightarrow$ probability inter-conversion is summarized by: 
\begin{equation}
0 \le s_{\text{p}} \equiv k \ln \left[ w_{\text{p}} \right] \equiv k \ln \left[ \frac{1}{p}\right] \le \infty
\label{surprisal}
\end{equation}
where of course the units are bits if $k = 1/\ln[2]$.

\subsection{average surprisals}
\label{sec:avgsurprisals}

The treatments of the ideal gas law, equipartition, mass action, and the laws of thermodynamics in the previous section connect to this tradition by defining uncertainty or entropy S as an {\bf average surprisal} e.g. in J/K between 0 and $+\infty$, Boltzmann's multiplicity W between 1 and $+\infty$ as $e^{S/k}$ where k is Boltzmann's constant, and 1/W as a reciprocal-multiplicity between 0 and 1. Their relevance to the thermal side of physics education has been discussed above.

More generally the interconversion for the average surprisal, uncertainty, or entropy associated with predicted probability-set $q$, as measured by operating probability-set $p$, can be written: 
\begin{equation}
0 \le S_{\text{p/p}} \le S_{\text{q/p}}  \equiv k \ln \left[ W_{\text{q/p}} \right] \equiv k \sum_{i=1}^N p_i \ln \left[ \frac{1}{q_i} \right]  \le \infty .
\label{uncertainty}
\end{equation}
Thus cross-entropy $S_{\text{q/p}}$ for an observation (in bits) is the {\bf average-surprisal when a proposed-model probability-set q differs from the operating-model probability-set p}. Although written for a discrete probability-set, the expression is naturally adapted to continuous as well as quantum-mechanical (amplitude-squared) probability-sets\cite{Jaynes57b}. 

Note that the upper limit on $S_{\text{p/p}}$ is $\ln_2[N]$. Also the fact that the cross-entropy $S_{\text{q/p}} \le S_{\text{p/p}}$, i.e. that {\bf measurements using the wrong model $q$ are always likely to be more surprised by observational data than those using the operating-model $p$}, underlies maximum-likelyhood curve-fitting and Bayesian model-selection as well as the positivity of the correlation and thermodynamic availability measures discussed below. 

Thus in this two-distribution case, $1 \le W_{\text{q/p}} \le +\infty$ is an {\bf effective choice-multiplicity} for expected probability set q in the face of operating-probability set p.  In general $W_{\text{p/p}} \le W_{\text{q/p}}$.  For the uniform N-probability set $u_i = 1/N$ for $i$ running from 1 to N, we can also say that $W_{\text{p/p}} \le W_{\text{u/p}} = N = W_{\text{u/u}} \le W_{\text{p/u}}$.

\subsection{net surprisals}
\label{sec:netsurprisals}

\begin{figure}
\includegraphics[scale=0.65]{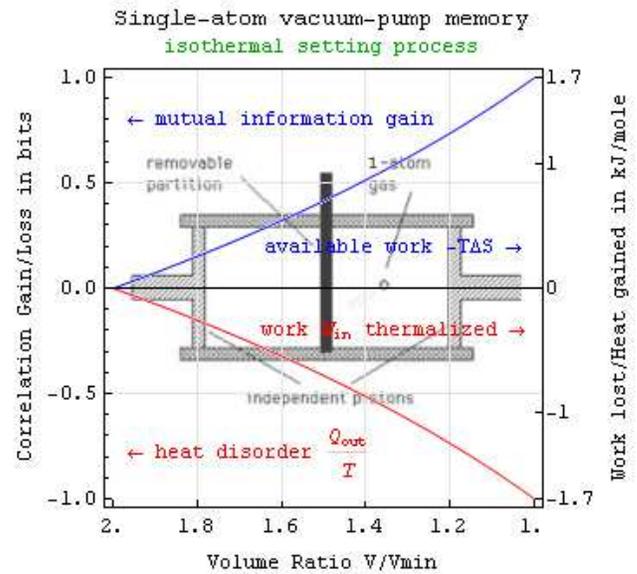}%
\caption{Szilard vacuum-pump memory schematic relating subsystem correlations to reversibly-thermalized work.}
\label{vacPumpMemory}
\end{figure}

\begin{figure*}
\includegraphics[scale=1.0]{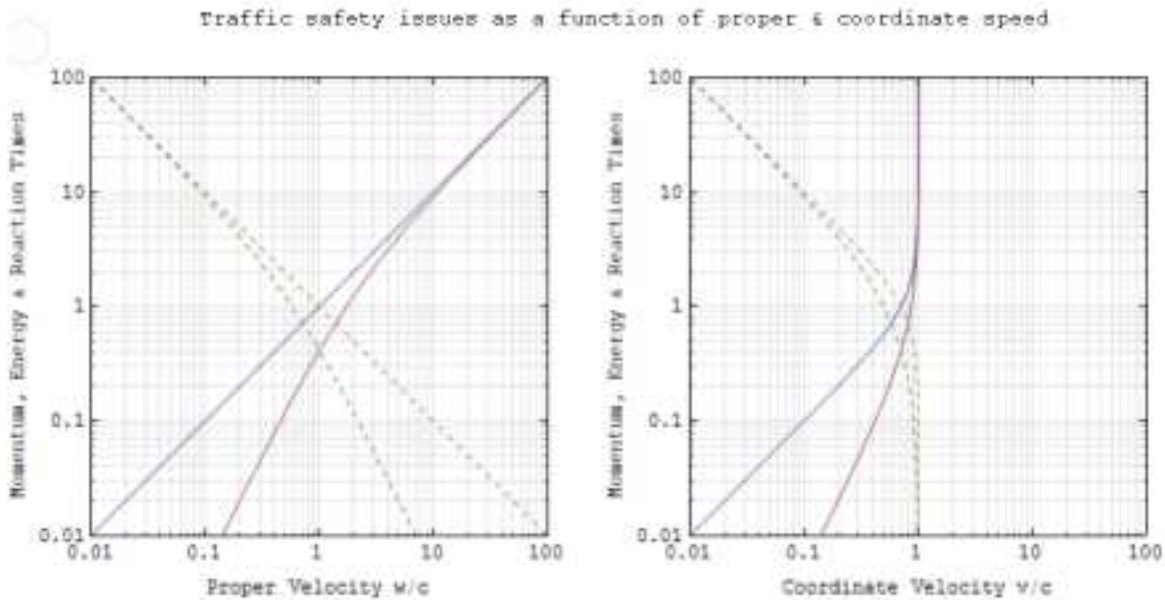}%
\caption{Momentum (blue), energy (red), \& driver/pedestrian (yellow/green) reaction-times vs. proper \& coordinate velocities.}
\label{figA1}
\end{figure*}

The tracking of subsystem correlations has taken a back seat in traditional thermodynamic use of log-probability measures. This is illustrated e.g. by the traditional treatment of subsystem entropies as additive, in effect promising that correlations between e.g. between gas atoms in two volumes separated by a barrier can be safely ignored. More generally, however, subsystem correlations of the form $I_{\text{AB}} = S_{\text{A}} + S_{\text{B}} - S_{\text{AB}}$ e.g. between a sent and a received message, or between traits of a parent and of a child are both non-zero and of central importance. Fortunately the maximum entropy discussed in macroscopic thermodynamics is little more than minimum KL-divergence with an uncorrelated prior\cite{Gregory2005}, so that physicists expert in its application to analog systems can already play a pivotal role informing students who take physics courses about these connections across disciplines. 

In particular the foregoing are backdrop to the paradigm-shift which broke out of physics into the wide world of statistical inference in the mid-20th century\cite{ShannonWeaver49}. We'll touch on only three of the many areas that it's connecting together today, based on their relevance to cross-disciplinary interests of students in physics classes. The specific application areas are: (i) {\bf thermodynamic availability} as in Appendix {\ref{appxF}, (ii) algorithmic {\bf model selection} as in Appendix {\ref{appxG}, and (iii) the {\bf evolution of complexity} as in Appendix {\ref{appxH}. The surprisal $\Leftrightarrow$ multiplicity $\Leftrightarrow$ probability interconversion for these correlation analyses may be written: 

\begin{equation}
0 \le I_{\text{q/p}} \equiv k \ln \left[ M_{\text{q/p}} \right]  \equiv k \sum_{i=1}^N p_i \ln \left[ \frac{p_i}{q_i} \right] \le \infty
\label{mutualinformation}
\end{equation}

Log-probability measures are useful for tracking subsystem-correlations in digital as well in analog complex systems. In particular tools based on Kullback-Leibler divergence $I_{\text{q/p}} \ge 0$ (the negative of Shannon-Jaynes entropy\cite{Jaynes1963,Gregory2005}) and the matchup-multiplicity or choice-reduction-factor $M_{\text{q/p}}$ associated with reference probability-set $q$ have proven useful: (i) to engineers for measuring available-work or {\em exergy} in thermodynamic systems\cite{Tribus71}, (ii) to communication scientists and geneticists for studies of: regulatory-protein binding-site structure\cite{Stormo98}, relatedness\cite{Bennett2003}, network structure, \& replication fidelity\cite{Cover2006,Barabasi2003}, and (iii) to behavioral ecologists wanting to select from a set of simple-models the one which is least surprised by experimental data\cite{Burnham2001,Burnham2002} from a complex-reality. 

Thanks to their experience with use of log-probability measures in analog physical-systems, physicists (plus non-physics students who take courses from physicists) can play a key role in the cross-disciplinary application of informatics to complex systems. These correlation-measures have 2nd law teeth (cf. \ref{appxH}), e.g. making them relevant to quantum computing\cite{Lloyd89b}, and they enable one to distinguish pair from higher-order correlations making them relevant to the exploration of order-emergence in a wide range of biological systems\cite{Schneidman2003,Harper2009}. They may be especially useful in addressing challenges associated with the sustainability of layered complex-systems, as discussed in Appendix \ref{appxJ}.  

\section{Discussion}
\label{sec:discuss}

The foregoing illustrates a simple qualitative-strategy for evaluating conceptual approaches to a given subject, inspired by quantitative developments across-disciplines in statistical inference. Putting together your own surprisal tables, in the areas discussed above as well as in others might be helpful (e.g. moels of continental-drift or of gene-based evolution) in clarifying why certain idea-sets are used now, and why they might change in the future. It may also help to share such tables with students and invite their input, since the perspective of experts wiii likely differ from the perspective of beginners particularly on the question of algorithmic complexity. 

Similar adhoc analyses of surprisal might also, for example, help {\em each of us} decide when it is (and is not) appropriate to spend time in the educational arena e.g. on: (i) geometric-algebra approaches\cite{Hestenes2003a,Hestenes2003b,Doran2003} to complex numbers \& cross-products, (ii) energy\cite{Moore2003} \& least-action\cite{Moore2004} based introductions to mechanics, (iii) vector potential introductions to magnetism\cite{Konopinski78}, (iv) explore-all-paths introductions to quantum mechanics\cite{Taylor98b}. The approach may even come in handy for sorting out differences in research strategy, e.g. in deciding how much time to spend (in context of a particular problem) on: (a) CPT approaches to the application of non-Hermitian Hamiltonians\cite{Bender2007}, (b) molecule-code as distinct from kin-selection models of evolving eusocial or altruistic behavior\cite{Nowak2010}, etc. 

\begin{acknowledgments}
For support and inspiration I would like to thank the 
late Bill Shurcliff and Ed Jaynes, not to mention a 
longer list of still-active colleagues, plus grad students 
Bob Collins, Zak Jost and Pat Sheehan for their 
participation in a Bayesian informatics development 
course that helped nail down some of the inter-connections 
discussed here.
\end{acknowledgments}


\appendix
\section{Proper-velocity}

Minkowski's flatspace metric-equation naturally defines the relation between traveler-time elapsed ($\delta \tau$) and the distance/time between events defined with respect to the yardsticks ($\delta \vec{r}$) and synchronized-clocks ($\delta t$) of a single map-frame. It thereby defines the interconversion between three ways to describe rate of travel, namely coordinate-velocity $\vec{v} \equiv \delta \vec{r}/d t$, proper-velocity $\vec{w} \equiv d\vec{r}/d \tau$, and the scalar Lorentz-factor $\gamma \equiv dt/d\tau$.  

Proper-velocity, referred to by Shurcliff as the ``minimally-variant'' parameter for describing position's rate of change, can simplify our understanding of many relativistic processes.  We choose three ways here that are relevant to an introductory physics course, the first two of which addresses the extent to which either coordinate or proper velocity best extrapolates familiar low-speed strategies to analysis at any speed.

\subsection{traffic safety}
\label{appxA}

\begin{center}
{\em If lightspeed were 55 [mph], would} {\bf you}

{\em take down the highway speed-limit signs?}
\end{center}

At speeds low compared to lightspeed $c$, the maximum possible collision-damage generally depends on vehicle kinetic-energy and momentum while the ability to minimize or avoid collision-damage depends on the reaction-times available to driver and pedestrian. These four quantities are graphed in the figure at right as a function of proper-velocity (on the left) and coordinate-velocity (on the right).

As you can see, momentum $p/mc = w/c$ scales linearly with proper-velocity $w/c$ at any speed while momentum's sensitivity to coordinate-velocity goes through the roof. Kinetic-energy $K/mc^2 = \gamma -1$ where $\gamma = \sqrt{1+(w/c)^2}$ likewise scales nicely with proper-velocity, although a slope-change in the super-relativistic limit brings kinetic-energy and momentum into increasingly good agreement.

In a complementary way, driver reaction-time $c d\tau/dx = c/w$ decreases inversely as proper-velocity at any speed while its sensitivity to coordinate velocity blows up. Pedestrian time-for-reaction after the warning photon arrives i.e. $c(dt/dx-1/c) = c\gamma/w-1$ likewise scales nicely with proper-velocity, but in this case a slope-change in the super-relativistic limit makes it inversely-proportional to proper-velocity squared.

Thus if one wishes to consider traffic-safety at high speeds, or in Mr. Tompkins style universes\cite{Gamow45} with much lower values for lightspeed, speed-limits should likely be expressed in units of proper and not coordinate velocity. In other words, limiting travelers to less than 55 map-lightyears per traveler-year makes more sense than limiting them to less than 0.999835 map-lightyears per map-year, especially since the former only requires that the speedometer divide traveler time into mile-marker (i.e. map) distance.

An added shortcoming of coordinate-velocity is that parameterizations in terms of it seldom lead to discussion of the super-relativistic limit on the high side of the transition at $w/c = 1$ map-lightyear/traveler-year. This transition between sub and super relativistic is seen in the figure at right as the 45 degree mark in the circular-tradeoff of motion-through-time ($d \tau /dt$) for motion-through-space ($dx/dt$). In addition to preserving low-speed properties, proper-velocity thus also leads to clearer views of the super-relativistic regime.

\subsection{3-vector addition}
\label{appxB}

\begin{center}
{\em Can familiar rules for adding 3-vector} 

{\em velocities be put to use at any speed?}
\end{center}

\begin{figure}
\includegraphics[scale=0.7]{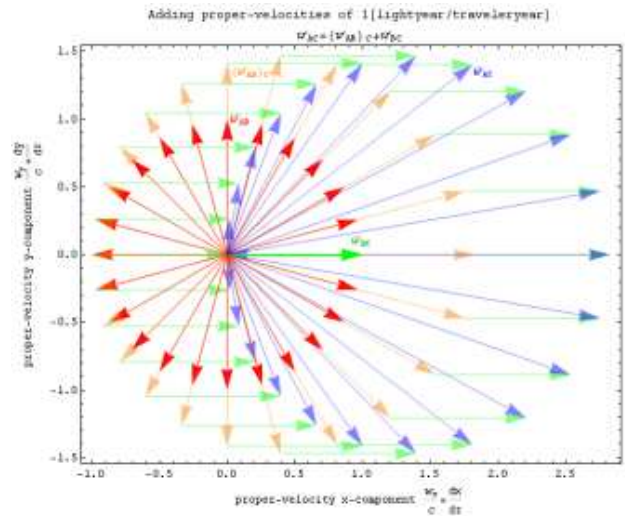}%
\caption{Adding $w/c=1$ proper-velocity vectors.}
\label{figB1}
\end{figure}

A useful mnemonic for relative motion in the Newtonian world is: 
\begin{equation}
\vec{v}_{AC} = \vec{v}_{AB} + \vec{v}_{BC}
\label{lospeed addition}
\end{equation}
where e.g. $\vec{v}_{AB}$ is the vector velocity of object $A$ with respect to object $B$.  Note that in general $\vec{v}_{AB} = -\vec{v}_{BA}$, and in sums ``a common middle letter cancels out''.

For a relation that works for {\em uni-directional} velocity-addition even at coordinate-speeds $v$ near lightspeed $c$, one might use the similar relationship:
\begin{equation}
w_{AC} \equiv \gamma_{AC} v_{AC} = \gamma_{AB} \gamma_{BC} \left( v_{AB} + v_{BC}\right)
\label{unidirectional}
\end{equation}
where the proper-velocity $\vec{w} \equiv d\vec{x}/d\tau = \vec{p}/m$ is map-distance $\vec{x}$ traveled per unit time $\tau$ on traveler-clocks, coordinate-velocity $\vec{v} \equiv d\vec{x}/dt$ with $v \le c$ is the usual map-distance traveled per unit time $t$ on map-clocks, and Lorentz-factor $\gamma \equiv dt/d\tau = 1/\sqrt{1-(v/c)^2} = E/mc^2 \ge 1$ is the ``speed of map-time per unit traveler-time'' provided that {\em the map-frame defines simultaneity}.  

Thus uni-directional coordinate-velocities add but Lorentz-factors multiply when forming the proper-velocity sum.  This allows colliders (sometimes with Lorentz-factors well over $10^5$) to explore {\bf much} higher-speed collisions than would be possible with a fixed-target accelerator. 

In the any-speed and {\em any-direction} case, neither this nor the familiar rule for adding coordinate-velocites takes the shape of 3-vector addition. However addition of proper-velocity 3-vectors can be written as:
\begin{equation}
\vec{w}_{AC} \equiv \gamma_{AC} \vec{v}_{AC} = \left(\vec{w}_{AB}\right)_{C} + \vec{w}_{BC}
\label{multidirectional}
\end{equation}
where C's view of the out-of-frame proper-velocity $(\vec{w}_{AB})_C$ is in the same direction as $\vec{w}_{AB} \equiv \gamma_{AB}\vec{v}_{AB}$ but rescaled (in magnitude only) by a factor of $(\gamma_{BC} + \vec{w}_{AB} \cdot \vec{w}_{BC}/(c^2(1+\gamma_{AB}))) \ge 0$, as illustrated for ``unit proper velocities'' in Fig. \ref{figA1}.  Hence the original low-speed equation {\em generalizes nicely} when proper-velocity $\vec{w} \equiv d\vec{x}/d\tau$ is used instead of coordinate-velocity $\vec{v} \equiv d\vec{x}/dt$.

\subsection{energy vs. momentum}
\label{appxC}

\begin{center}
{\em What does relativity say about} {\bf everything}

{\em on a kinetic-energy versus momentum plot?}
\end{center}

\begin{figure}
\includegraphics[scale=0.7]{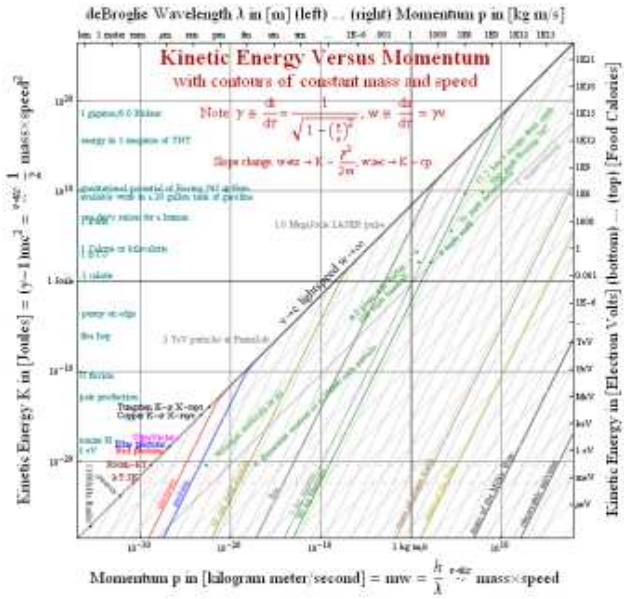}%
\caption{Kinetic energy vs. momentum plot.}
\label{figC1}
\end{figure}

Vector proper-velocity $\vec{w}$ like vector momentum $\vec{p} = m\vec{w} = m \gamma \vec{v}$ has no size upper-limit.  Likewise for scalar Lorentz-factor $\gamma$, which like kinetic energy $K = (\gamma -1) m c^2$ has no intrinsic upper-limit.  

Hence Newton likely imagined that by choosing the appropriate mass $m$, objects may be found with any desired mix of kinetic energy $K$ and translational momentum $p$.  As shown in Figure \ref{figB1}, however, Minkowski's metric equation:
\begin{equation}
(c \delta\tau)^2 = (c \delta t)^2 - \vec{\delta r} \cdot \vec{\delta r}
\label{metriceqn}
\end{equation}
by defining Lorentz-factor in terms of coordinate-velocity in effect lowers a curtain on kinetic-energy/momentum space by making only the lower right half of it accessible to moving objects.

The log-log plot in Figure \ref{figB1}, which also has lines of constant mass and constant coordinate-velocity, thus provides students with an integrative view of kinetic-energy and momentum space for a wide range of objects in (and beyond) everyday experience.  Thus for example if one points these relationships out as early as possible in an intro-physics course (perhaps as early as the kinematics-section on relative velocities if one takes the time to distinguish traveler-time $\tau$ from map-time $t$ in defining Lorentz factor $\gamma \equiv dt/d\tau$ as in the previous section), then one may find opportunities again-and-again to refer back to it as new phenomena come up in the course.

\section{Proper-acceleration}

By sticking with a single map-frame to define extended simultaneity, one finds that equations for accelerated motion also extapolate nicely from low to high speed.  The organizing parameter for this extension is the three non-zero (spatial) components of an object's acceleration four-vector as seen from the vantage point of the object itself.  

This 3-vector is referred to as the object's proper-acceleration $\vec{\alpha}$.  Again we discuss two uses for this quantity that are most relevant to students in an intro-physics course.  

\subsection{proper or geometric?}
\label{appxD}

\begin{center}
{\em Equivalence replaces real versus inertial forces}

{\em with proper versus geometric.}
\end{center}

For intro-physics students even at low speeds one might point out that there are experimentally two kinds of acceleration: proper-accelerations associated with the push/pull of external forces, and geometric-accelerations caused by choice of a reference-frame that is not geodesic i.e. a local reference coordinate-system that is not``in free-float''.

Typically proper-accelerations are felt through their points of action e.g. through forces on the bottom of your feet, or through interaction with electromagnetic fields. On the other hand, geometric-accelerations associated with one's coordinate choice are associated with affine-connection forces (an extended version of the Newtonian concept of inertial force) that {\em act on every ounce of an object's being}. 

Affine-connection effects either vanish when seen from the vantage point of a local free-float or geodesic frame (an extended version of the Newtonian concept of inertial frame), or give rise to non-local force effects on your mass distribution which cannot be made to disappear.  Some of these are summarized in Table \ref{tableC1}. 

\begin{table}
\caption{Acceleration types and various forces.}
\begin{tabular}{r|c|c|c}
force name type 			& proper	& geometric: non-free	& non-local\\ 
\hline \hline
normal 						& $\oplus$  			&	& 		\\
\hline
string 						& $\oplus$ 				&	& 			\\
\hline
spring 						& $\oplus$ 				&	& 			\\
\hline
friction 						& $\oplus$ 				&	& 			\\
\hline
drag 						& $\oplus$ 				&	& 			\\
\hline
centripetal 						& $\oplus$ 				&	& 			\\
\hline
electromagnetic 						& $\oplus$ 				&	& 			\\
\hline
gravity 						&	& $\oplus$ 	& 			\\
\hline
reaction ``gees" 						&	& $\oplus$ 	& 			\\
\hline
centrifugal 						&	& $\oplus$ 	& 			\\
\hline
Coriolis effect 						&	&	& $\oplus$  			\\
\hline
tidal 						&	&	& $\oplus$  			\\
\hline \hline
\end{tabular}
\label{tableC1}
\end{table}

Although the following need not be shared, the assertion above contains the essence of general relativity's {\bf equivalence principle} which guarantees that Newton's Laws can be helpful {\em locally} in accelerated frames and curved space time, provided that we invoke {\bf inertial forces} to explain the geometric-accelerations which operate in those frames.

The mathematics of geometric accelerations comes from the fact that in {\em general relativity} an object's coordinate acceleration (as distinct from only its proper-acceleration 4-vector $A$) is equal to:
\begin{equation}
\frac{dU^\lambda }{d\tau } =A^\lambda - \Gamma^\lambda {}_{\mu \nu}U^\mu U^\nu
\end{equation}
where geometric-accelerations are represented by the affine-connection term $\Gamma$ on the right hand side.  These may be the sum of as many as sixteen separate velocity and position dependent terms.  Coordinate acceleration goes to zero whenever proper-acceleration is exactly canceled by that connection term, and thus when physical and inertial forces add to zero.

\subsection{accelerated roundtrips}
\label{appxE}

\begin{center}
{\em Metric-1st approaches make accelerated-twin}

{\em roundtrips easy to analyze at any speed.}
\end{center}

\begin{figure}
\includegraphics[scale=0.7]{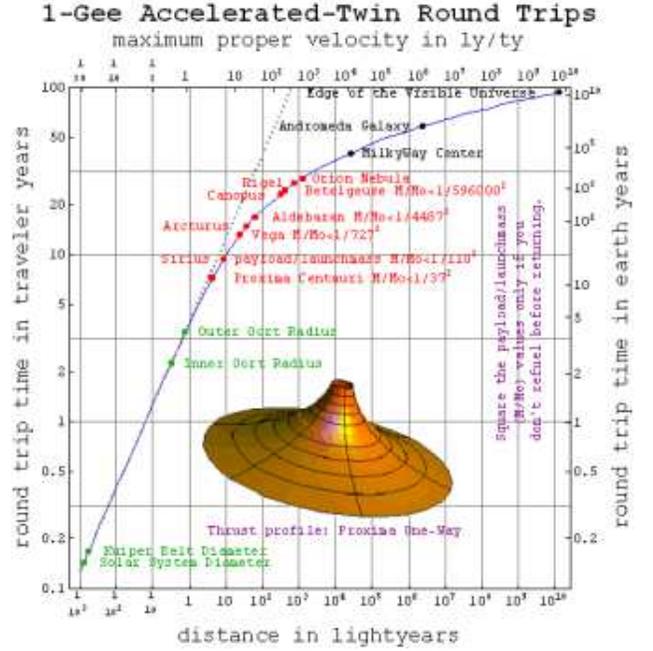}%
\caption{1-gee proper-acceleration roundtrips.}
\label{figD1}
\end{figure}

For unidirectional (1+1)D motion, the rapidity or hyperbolic velocity angle $\eta$ simply connects the interchangable velocity parameters Lorentz-factor $\gamma \equiv dt/d\tau$, proper-velocity $w \equiv dx/d\tau$ and coordinate-velocity $v \equiv dx/dt$ via:

\begin{equation}
\eta \equiv \sinh^{-1}\left[\frac{w}{c}\right] = \tanh^{-1}\left[\frac{v}{c}\right] = \pm \cosh^{-1}\left[\gamma\right]
\end{equation}

These parameters may then be used to express the proper-acceleration $\alpha$ experienced by an object traveling with respect to a map-frame of co-moving yardsticks and synchronized clocks in flat space time, in terms of its coordinate-acceleration a which cannot be held constant at high speed, as:

\begin{equation}
\alpha \equiv \frac{1}{\gamma} \frac{dw}{d\tau} = \gamma^3 a  \text{, where } a \equiv \frac{dv}{dt}
\end{equation}

This yields three integrals of constant proper-accelerated motion that reduce to the familiar equations of constant coordinate-acceleration at low speeds:

\begin{equation}
\alpha = \frac{\Delta w}{\Delta t} = c \frac{\Delta \eta}{\Delta \tau} = c^2 \frac{\Delta \gamma}{\Delta x} \stackrel{v\ll c}{=} \frac{\Delta v}{\Delta t} \stackrel{v\ll c}{=} \frac{1}{2} \frac{\Delta (v^2)}{\Delta x}
\end{equation}

These in turn allow one for example to write out analytical solutions (cf. Fig. \ref{figD1}) for round-trips involving constant 1 gee $\simeq 1.03[\text{ly}/\text{y}^2]$ accelerated/decelerated travel between stars.

\section{Choice multiplicities}

Senior physics courses have for already been re-arranged considering the fundamental role that multiplicity (and its logarithm, namely entropy) play in understanding and predicting behaviors.  Although intro-physics courses are weaker in this context, books like Tom Moore's ``Six Ideas''\cite{Moore2003} have put choice-multiplicity where it belongs at the start of the thermo-chapters. 

Hence the only section in this Appendix is one for students with virtually no math background. The hope is that teachers will individually explore ways to introduce the connection between bits and J/K, while at the same time nurturing an appetite for textbook revisions that better communicate the relation between thermal physics and information theory downstream.

\subsection{quantifying risk}
\label{appxF}

\begin{center}
{\em N bits of surprisal (i.e. N heads in N tosses)}

{\em attaches modest numbers to very tiny risks.}
\end{center}

\begin{figure}
\includegraphics[scale=1.0]{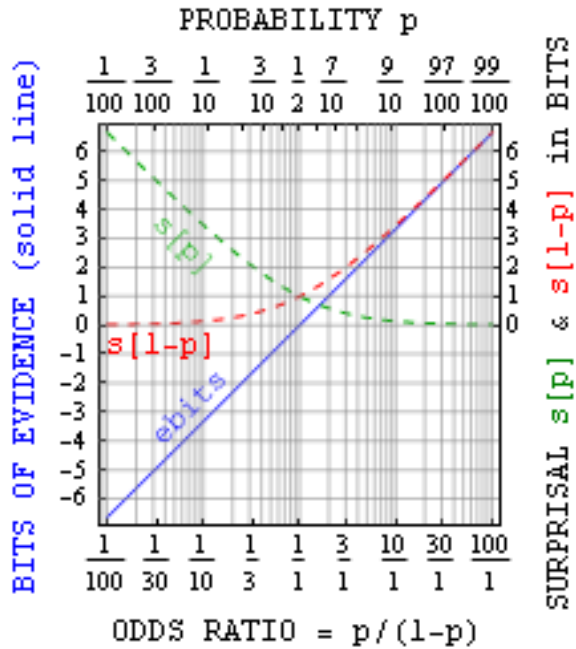}%
\caption{Logarithmic measures of probability and odds.}
\label{figE1}
\end{figure}

Surprisal in bits (defined by probability = $1/2^{\#\text{bits}}$) might be useful to citizens in assessing risk and/or standards of evidence (cf. Fig. \ref{figE1}), because of its simple, intuitive, and testable ability to connect even very small probabilities with one's experience at tossing coins. For example, the surprisal of dying from a smallpox vaccination (one in a million) is about 19.9 bits (like 20 heads in 20 tosses), while the surprisal of dying from smallpox once you have it (one in three) is only about 1.6 bits (i.e. more likely than 2 heads in 2 tosses). 

Thus surprisal: (i) has meaning which is easy to remind yourself of with a few coins in your pocket, (ii) reduces huge numbers to much more intuitive size, and (iii) allows one to combine risks ''from independent events'' with addition/subtraction rather than multiplication/division. 

For instance (from the numbers suggested above) your chance of dying is decreased by getting the vaccination, as long as the surprisal of getting smallpox without the vaccination is less than $20 - 2 \simeq 18$ bits. That means that vaccination is your best bet (absent other information) if your chances of being exposed to smallpox are greater than those of getting 18 heads in 18 tosses (1 out of $2^{18} \simeq 333,333$).

Given the large difference between something with 2 bits of surprisal and something with 18, communications bandwidth might be better spent by newsmedia providing us with numbers on observed surprisal, rather than by reporting only that ``there's a chance'' of something bad (or good) happening. Saying the latter treats your audience as consumers of spin rather than information. 

Likewise, use of surprisals in communicating and monitoring risks to medical patients could make patient decisions about actions with a small chance of dire outcomes as informed as possible. This could reduce the costs of medical malpractice in the long run by empowering patients with tools to make informed and responsible choices, making the need for legal redress less frequent.  

Thus the media for risk-assessing public could play a key role in reducing the costs of defensive medicine. Some might even enjoy surprisal data on the small probabilities associated with some gambling opportunities.  After all, there really is more to the lottery than simply knowing ``the size of the pot''.

\subsection{freedom of choice}
\label{appxG}

\begin{center}
{\em How is the number of choices affected}
 
{\em by their likelihood of being chosen?}
\end{center}

Just as $\#_{\text{choices}} = 2^{\#_{\text{bits}}}$ is both a practical and accessible way to introduce information units and surprisals to students with or without a lot of math background, so the {\em effective} number of accessible states $W = e^{S/k}$ is a useful way to discuss entropies and average surprisals.

This has applications in thermal physics since when extensive work-parameters like volume, energy, and number of particles are fixed (Gibbs' micro-canonical case), all states may be treated as equally probable and lovely expressions follow for the assumptions behind (and hence application domains for) the ideal gas law, equipartition, and mass-action. Although simply counting states is nice, the information we have available often results in choices that are not all equally probable.

For example when a system is in contact with a thermal reservoir (Gibbs' canonical ensemble) the average energy is fixed by the energy uncertainty-slope $dS/dE)_{VN} = 1/kT$. Then $W$ becomes an effective number of accessible states, each of whose occupation-probability depends on its thermal energy $\varepsilon$ in proportion to the Boltzmann-factor $e^{-\varepsilon /kTk}$.

Moving back from physics to statistical inference, therefore, knowledge of occupation probabilities $p_i$ for a set of states $i=1,N$ allows one to calculate both uncertainties (in information units) and the associated multiplicity (i.e. an effective measure of number of choices). In the notation of this paper, that is: 

\begin{equation}
1 \le W_{\text{p/p}} = \prod_{i=1}^{N}\left(\frac{1}{p_{i}}\right)^{p_{i}} = 2^{\#_{\text{bits}}} \le N
\label{Wpp}
\end{equation}

This quantity has some interesting uses outside of physics. It is important in data-compression, for example, since a message with letter-patterns seldom used may be more compactly communicated in a language with all symbols of equal likelihood\cite{ShannonWeaver49,Cover2006}. It might also help analyze the effective number of food-choices in a restaurant if customers mainly order one item even though 10 are on the menu, particularly if you don't want something that has been sitting unrequested for week.

The nice thing about pedagogical use of choice-multiplicities (as long as the numbers are small) is that they may be less mysterious than entropy, even though both have a rather messy definition in terms of state probabilities. Another possible application of this will be mentioned in Appendix \ref{appxJ}.

\section{Subsystem correlations}

Of the integrative concepts discussed in this paper, the least familiar to physicists (judging from textbooks, at least) may be those associated with the logarithmic correlation-measure sometimes referred to as Kullback-Leibler divergence and its multiplicity, a kind of normalized choice-reduction factor which is never less than 1. Hence in this Appendix we discuss the connection of these measures to: (i) available work, (ii) information engines relevant e.g. to energy-flow in both biology and in computer science, and (iii) the evolution of {\em multi-layer} complex analog as well as digital systems. 

\subsection{work's availability}
\label{appxH}

\begin{center}
{\em What's the highest ambient T for}

{\em unpowered-conversion of 100C to 0C water?}
\end{center}

\begin{figure}
\includegraphics[scale=0.7]{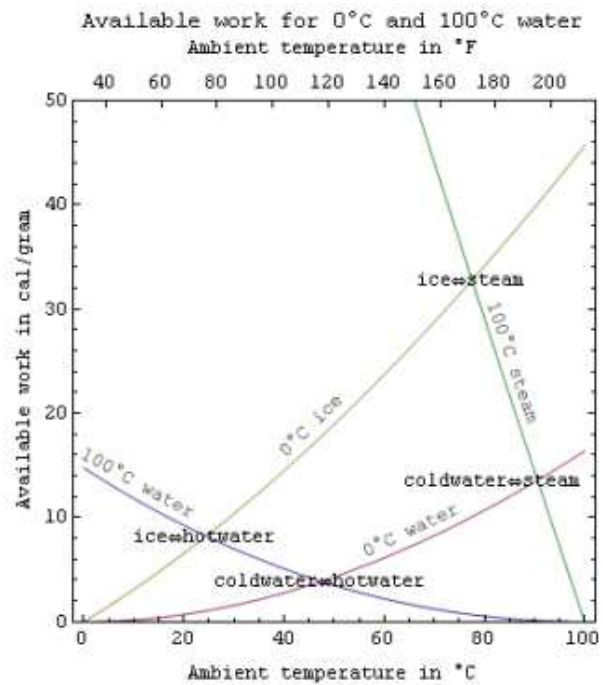}%
\caption{Available work/gram vs. ambient temperature $T_o$ for $0^\circ$ and $100^\circ$ water in various states.}
\label{figF1}
\end{figure}

Best-guess states (e.g. for atoms in a gas) are inferred by maximizing the average-surprisal S (entropy) for a given set of control parameters (like pressure $P$ or volume $V$). This constrained entropy maximization, both classically and quantum mechanically, minimizes Gibbs availability in entropy units $A = -k \ln[Z]$ where $Z$ is a constrained multiplicity or partition function.

When absolute temperature $T$ is fixed, free-energy ($T$ times $A$) is also minimized. Thus if $T$, $V$ and number of molecules $N$ are constant, the Helmholtz free energy $F = U - T S$ (where $U$ is energy) is minimized as a system ``equilibrates". If $T$ and $P$ are held constant (say during processes in your body), the Gibbs free energy $G = U + P V - T S$ is minimized instead. The change in free energy under these conditions is a measure of available work that might be done in the process. Thus available work for an ideal gas at constant temperature $T_o$ and pressure $P_o$ is $W = \Delta G = N k T_o \Theta [V/V_o]$ where $V_o = N k T_o / P_o$ and by Gibbs inequality $\Theta [x] \equiv x-1- \ln[x] \ge 0$.

When this Gibbs function is applied to a ratio between subsystem and ambient parameter-values which have a power-law relation to state multiplicity $W$ (as do energy in quadratic systems, and volume in ideal gases), the function may be used to express the entropy gained by the ambient minus that lost by a subsystem allowed to equlibrate with that ambient. This entropy-difference equals the KL-divergence or net-surprisal $\Delta I \ge 0$, defined as the average value of $k \ln[p/p_o]$ where $p_o$ is the probability of a given state under ambient conditions. KL-divergence times ambient temperature $T_o$ therefore yields the work available on thermalization.

For instance, the work available in equilibrating a monatomic ideal gas to ambient values of $V_o$ and $T_o$ is thus $W = T_o \Delta I$, where KL-divergence $\Delta I = N k (\Theta[V/V_o]+(3/2)\Theta[T/T_o])$. The resulting contours of constant KL-divergence put limits on the conversion of hot to cold as in flame-powered air-conditioning, or in the unpowered device to convert boiling-water to ice-water (Fig. \ref{figF1}). 

\subsection{information engines}
\label{appxI}

\begin{center}
{\em Life depends on subsystem-correlations created}

{\em by reversibly-thermalizing available-work.}
\end{center}

\begin{figure}
\includegraphics[scale=0.57]{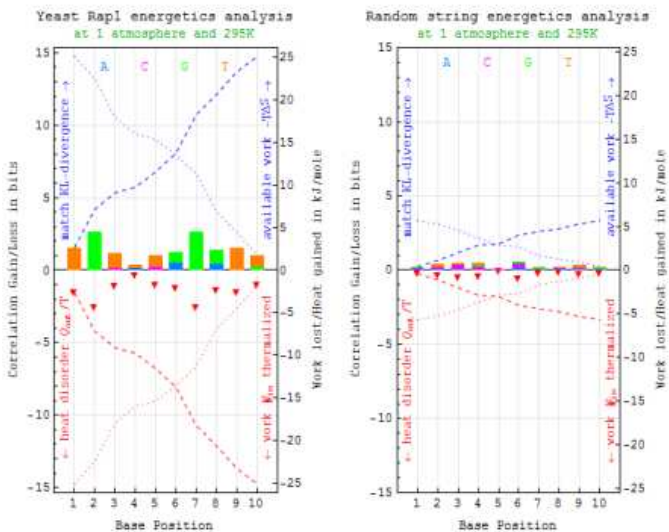}%
\caption{Yeast Rap1 versus equiprobable sequence-energetics, relating correlations to work as in Fig. \ref{vacPumpMemory}.}
\label{figI1}
\end{figure}

As mentioned in the previous appendix, dimensionless KL-divergence (first suggested by Gibbs\cite{Gibb1873}, and known to engineers as {\em exergy} in units of $k T_o$) is a parameter-independent measure of thermodynamic availability. For changes which obey ensemble-constraints (constant $T$ in the canonical ensemble case) it may also obtained by dividing free-energy changes by $k T_{\text{ambient}}$. Reversible-thermalization occurs (e.g. in Fig. 2) when Gibbs-availability is converted from an ordered-energy (work) form into a non-energy form (like subsystem-correlations) while the energy associated with that input work is dropped into the ambient reservoir as heat. 

One consequence of this (as earlier illustrated in Fig. \ref{vacPumpMemory}) is that a nanoJoule of available work in a room-temperature ambient can be used to reversibly generate as much as $1/((273.15+22) k_B \ln[2])) \simeq 44.27$ gigabytes of subsystem KL-divergence. Lower temperature means higher uncertainty-slope. In fact natural-units for reciprocal-temperature or coldness (e.g. GB/nJ) directly specify the amount of subsystem-correlation that might by created per unit energy in such a process. Thus cooling a CCD can reduce background-noise, and irreversibly-thermalizing a gallon of gasoline without reason is a considerable waste even if we weren't trying to limit our contributions to atmospheric $C O_2$.

In spite of reversible-thermalization's counter-intuitive look in an engine-diagram (i.e. work is thermalized while the order associated with it is retained), it has long been of quantitative importance in the evolution of molecular codes. It also promises to be important in the future of computing. In the broader field of statistical inference, moreover, its applications go way beyond the world of atoms because KL-divergence is a comprehensive measure of subsystem-correlations, even when the observed-averages used as inputs do not represent quantities which might be conserved (like energy and volume) on random-exchange between systems. 

For example Tom Schneider at the {\em National Cancer Institute} and Gary D. Stormo at {\em Washington University} have explored the application of information theory to analysis of genetic-sequence variability \& evolution\cite{Schneider1986,Stormo1998a}.  The focus here is on the molecular mechanisms to inform protein-expression processes (which take place {\em inside} individual-cells) about processes-afoot in the organism and larger world around, something that the eukaryotic cells of plants and animals do especially well.

In a nutshell, regulatory proteins (sometimes called transcription-factors) bind to specific regions of the genomic DNA in order to influence the transcription of nearby genes.  The experimental surprise was that those binding sites that are key to natural selection often deviate from randomness by almost precisely the amount of information needed for molecular recognition.  

In other words it appears that for many types of DNA binding site, that site's KL-divergence or net-surprisal relative to ambient, e.g.
\begin{equation}
I_{\text{sequence}} =  \sum_i^L I_i \text{ where } I_i \simeq \sum_{\text{b}}^{\text{A,C,G,T}} f_{\text{bi}} \ln_2 \left[ \frac{f_{\text{bi}}}{p_b}\right]
\end{equation}
in bits, and hence the difference (in units of $R T \ln[2]$) between free-energy for binding to that sequence and the free-energy for binding to an average sequence\cite{Fields1997}, evolves to match the information in bits (where $\#_{\text{choices}} = 2^{\#_{\text{bits}}}$) needed to locate the site e.g.
\begin{equation}
I_{\text{freq}} \simeq \ln_2 \left[\frac{\# \text{bases}}{\# \text{sites}} \right]
\end{equation}
and no more.  This localizing information depends (to first order) on only the site frequency (i.e. binding sites per nucleotide), and hence only on the genome size ($\#$ bases) and the number of binding sites for a given protein on that genome ($\#$ sites).  Put another way, this is a situation on the nano-scale where code-string information content evolves to limits specified by the equilibrium thermodynamics of subsystem correlations, as shown in Fig. \ref{figI1} in much the same way as in Szilard's vacuum pump memory (Fig. \ref{vacPumpMemory}).

Exploring the physical processes by which these correlations evolve is a present-day cross-disciplinary challenge. Recent work at Princeton\cite{Schneidman2003} has shown that total correlation, a special case of KL-divergence and an extension of mutual information, can be dis-assembled with maximum-entropy techniques into pair and post-pair information terms. 

Pair correlations can be illustrated with a $2 \times 2$ kangaroo problem\cite{Gull1984}. For example if half of all kangaroos have blue eyes and also half of them are left-handed, estimate the fraction $p_{11}$ of kangaroos that are both blue-eyed and left-handed. Here $p_{11}$ runs from 0 to 1/2, with a uniform-prior (maximum-entropy) best-guess at $p_{11}$ = 1/4. Pair correlations over this interval run from 1 bit (blue always goes with right) through 0 at the best-guess point (color says nothing about handedness) back up to 1 bit (blue always goes with left), while distribution entropy runs from 1 bit (either right-blue or not) through 2 bits (left or right, blue or not) up to 1 bit (either left-blue or not). 

The simplest system to illustrate post-pair correlations may be a $ 2 \times 2 \times 2$ kangaroo system e.g. in which we know that a quarter of all kangaroos are female with blue eyes, a quarter are male with blue eyes, a quarter are female and left-handed, a quarter are male and left-handed, a quarter are left-handed with blue eyes, and a quarter are right-handed with blue eyes. The fraction $p_{111}$ of kangaroos that are left-handed blue-eyed females varies over its allowed range from 0 through 1/8 (best-guess) to 1/4. Pair correlations are zero for all possible choices (e.g. color alone says nothing about handedness), with the extreme values of $p_{111}$ showing exactly one bit of post-pair correlation (2 bits of uncertainty as one of 4 equally-likely choices), while the best guess shows no correlations at all (3 bits of uncertainty as one of 8 equally-likely choices). When $p_{111}$ is zero, of course, the 4 choices are brown-right-male and its 3 ``doubly-different'' permutations, while when $p_{111}$ is 1/4, the choices are blue-left-female etc. 

The Schneidman et al.\cite{Schneidman2003} strategy allows us to breakdown any experimentally-observed subsystem correlations into pair and post-pair components, making it natural to ask: Do subsystem correlations, e.g. between chemicals in a cell or between neurons associated with a sense-organ, first develop as pair correlations, then pair-pair correlations, etc. as the system becomes increasingly sophisticated? 

Preliminary reports\cite{Bialek2007} suggest that pair-correlations constitute the lion's share of the interaction even in highly evolved systems. If we want to understand the evolution of hierarchically-ordered complex systems, like planetary surfaces with bilayer-membrane enclosed cells organized into organisms that can work together, we may have to look to the thermodynamics of boundary-emergence and symmetry-breaking discussed briefly in Appendix D3.

\subsection{layered correlations}
\label{appxJ}

\begin{center}
{\em Correlations looking in/out from a layered-}

{\em hierarchy of subsystem-boundaries may evolve.}
\end{center}

\begin{figure}
\includegraphics[scale=0.7]{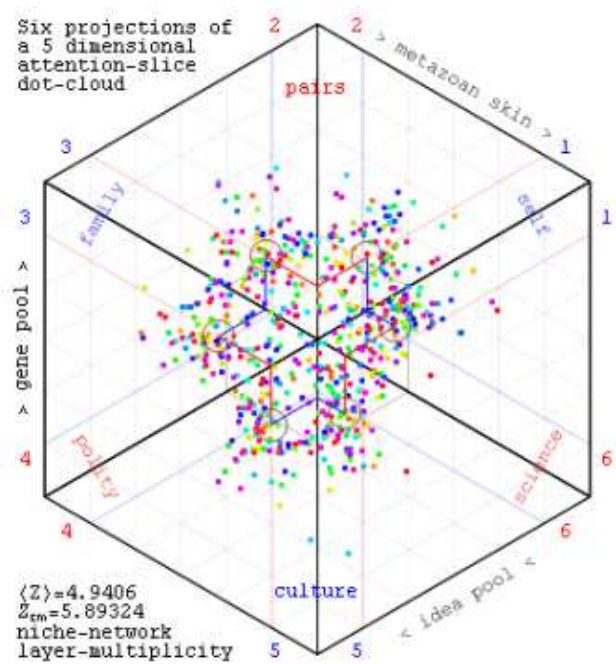}%
\caption{Each community member shows up as a dot, in each of the six triangular projections.}
\label{figH1}
\end{figure}

A cross-cutting thread in modern physics today is that of symmetry-breaking and boundary-emergence\cite{Anderson72}. It lies at the heart of quantum measurement and of vacuum cosmology, as well as our understanding of phase changes\cite{Sethna2006}, nucleation/precipitation, and the physical evolution of star/planetary systems\cite{Chaisson01}. 

One can see also its signature in the evolution of biogeochemical systems, powered partly by available-work from high-energy solar/stellar photons. In living systems, work is already underway to follow the first two steps in Chapter 9 of Sethna\cite{Sethna2006}, namely (i) identify broken symmetries associated with each layer of organization, and (ii) define order parameters to track their developments. In the process, we will also call attention to exciting and largely unexplored prospects for considering multiple layers of organization at the same time.

\subsubsection{emergent boundaries}

In the steady-state gradient setup between the earth's planetary surface and its solar-system environment, bilayer membranes appear to have made possible the symmetry-break between chemical-environments inside and outside of our earliest microbes. Inward looking (post-pair) molecule-correlations from the cell-perspective thus take on evolutionary meaning only when outward-looking cell-to-cell pair-correlations do the same. 

The replication of in-cell chemical correlations presumably at some point went from analog to digital with the evolution of the nucleic-acid/amino-acid system of molecular-codes that is now largely ubiquitous in life on earth. A similar transition from analog to digital idea codes, of course, is likely underway today. 

The much later development of multicelled lifeforms was presumably predicated by the development of organ surfaces, and of organism surface-boundaries like the metazoan skin. These in turn opened the door in parallel to the co-evolution of post-pair correlations between organ systems and pair-correlations between individuals. 

The physical boundaries that define metazoan communities beyond the skin are a bit more subtle. Perhaps the first of these boundaries emerged when a subset of multi-celled animals began to treat their young differently than the young of other members of the same species. With this break in symmetry, family gene-pools may have been born as evolving structures, providing new traction for the development of internal post-pair correlations between individuals, and external pair-correlations between families.

Most recently, as mentioned earlier, idea-codes to mediate correlated organism-behaviors have taken on a life of their own with the development of language, phonetic alphabet, printing, and most recently digital electronic communications. The symmetry-break between individuals who share your ideas, and those who don't, is likewise giving rise to the evolution of mechanisms for communication within and between cultures.

\subsubsection{order parameters}

The second task, of finding order parameters in these areas, has a long and cross-disciplinary history. However much work has concentrated on the buffering of correlations one level at a time. The correlation-first approach to complex-systems discussed here, therefore, might inspire students in physics classes to downstream help explore the utility of such order-parameters across multiple layers of organization.

For example, multi-celled animals (metazoans) actively assist in buffering correlations that look in and out from at least the last three boundaries discussed in the previous section i.e. skin, family and culture. The simplest possible multi-layer approach, therefore, might be to explore the layer choice-multiplicity for individuals in a given metazoan community. 

One might imagine, correctly or not for example, that certain spiders put a smaller fraction of their time and resources toward attending to family matters and between-family politics than do certain bears. If so the effective freedom-of-choice (or niche-network layer-multiplicity) for spiders in their communities may be smaller than the layer choice-multiplicity for bears in theirs. 

Moreover, following a natural disaster the task layer-multiplicity for any given community type might be disrupted further at least temporarily. In that sense task layer-multiplicity might be considered a measure of community health and, in a more limited way, adaptability.

If we have a way to estimate the fraction $f_{ji}$ of their effort that individual $j=1,N$ manages to put into layer $i=1,6$, where $\sum_i f_{ji} = 1$, then the building blocks of this order parameter are each individual's choice multiplicity $w_j$ (introduced in Appendix C2) and their cross-multiplicity $w^*_j$ with respect to the community average values $\langle{f_i\rangle} \equiv (1/N)\sum_j f_{ji}$:
\begin{equation}
w_j \equiv \prod_{i=1}^6 \left(\frac{1}{f_{ji}}\right)^{f_{ji}} \le \prod_{i=1}^6 \left(\frac{1}{\langle f_i\rangle}\right)^{f_{ji}} \equiv w^*_j.
\end{equation}
For the population as a whole, the geometric averages of these two quantities, namely:
\begin{equation}
W_{\text{char}} \equiv \prod_{j=1}^N \left(w_j\right)^{\frac{1}{N}} \le \prod_{j=1}^N \left(w^*_j\right)^{\frac{1}{N}} \equiv W_{\text{cm}},
\end{equation}
yield for the community a characteristic layer-multiplicity $W_{\text{char}}$ (corresponding to the average layer-uncertainty in log-probability space) and a center-of-mass multiplicity $W_{\text{cm}}$ (corresponding to the average cross-entropy with respect to the community average), which take on unit-less values between 1 and 6. The ratio between the two corresponds to an average KL-divergence, which speaks to niche-diversity within a given population.

By way of example, the community in the figure is drawn by random simplex-point picking to uniformly span all possible niche-network layer-assignments. Its center-of-mass multiplicity $W_{\text{cm}}$ is about 6, suggesting that all 6 correlation-layers are getting their share of attention. However its characteristic layer-multiplicity is $W_{\text{char}} \simeq 4.26$, suggesting that individuals in such communities may not typically be evolved to worry about six layers at once. This may be a scientific way to say why ``It takes a village.''

Data on attention-fractions e.g. from self-reporting and communications traffic might allow folks to track the state of existing communities e.g. as a function of environmental and policy changes, as well as to model layer-evolution. Thus we offer it here as something that students in physics classes might enjoy playing with, as well something that might inspire future physics contributions to important-problems in the study of complex systems.


\bibliography{ifzx2}

\end{document}